\documentclass{emulateapj}
\usepackage{natbib}

\shorttitle{A Massive Neutron Star in M5}
\shortauthors{Freire et al.}

\begin{document}

\title{A Massive Neutron Star in the Globular Cluster M5}

\author{Paulo C. C. Freire\altaffilmark{1},
Alex Wolszczan\altaffilmark{2},
Maureen van den Berg\altaffilmark{3},
Jason W. T. Hessels\altaffilmark{4}}
\altaffiltext{1}{N.A.I.C., Arecibo Observatory, HC 03 Box 53995, PR 00612,
  U.S.A.; {\tt pfreire@naic.edu}}
\altaffiltext{2}{Department of Astronomy and
  Astrophysics, Penn State University, University Park, PA 16802,
  U.S.A.; {\tt alex@astro.psu.edu}}
\altaffiltext{3}{Harvard-Smithsonian Center for Astrophysics, 60
Garden Street, Cambridge, MA 02138, U.S.A., {\tt
  maureen@head.cfa.harvard.edu}}
\altaffiltext{4}{Astronomical Institute ``Anton Pannekoek'',
University of Amsterdam, Kruislaan 403, 1098 SJ Amsterdam, The
Netherlands; {\tt jhessels@science.uva.nl}}

\begin{abstract}
We report the results of 19 years of Arecibo timing for two pulsars in
the globular cluster NGC~5904 (M5), PSR~B1516+02A (M5A) and
PSR~B1516+02B (M5B). This has resulted in the measurement of the
proper motions of these pulsars and, by extension, that of the cluster
itself. M5B is a 7.95-ms
pulsar in a binary system with a $> 0.13\,M_{\sun}$ companion and
an orbital period of 6.86 days. In deep HST images, no optical
counterpart is detected within $\sim 2.5\sigma$ of the position of the
pulsar, implying that
the companion is either a white dwarf or a low-mass main-sequence star. The
eccentricity of the orbit ($e = 0.14$) has allowed a measurement of
the rate of advance of periastron:
$\dot{\omega}\,=\,(0.0142\,\pm\,0.0007)^\circ \rm yr^{-1}$.
We argue that it is very likely that this periastron advance is due to
the effects of general relativity, the total mass of the binary system
then being $(2.29\,\pm\,0.17)\,M_{\sun}$. The small measured mass
function implies, in a statistical sense, that a very large fraction
of this total mass is contained in the pulsar: $M_p \, = \,
(2.08 \pm 0.19)\, M_{\sun}$ (1 $\sigma$); there is a 5\%
probability that the mass of this object is $<\,1.72\,M_{\sun}$
and a 0.77\% probability that $1.2 \,M_{\sun} \, \leq M_p \, \leq\,
1.44\,M_{\sun}$.  Confirmation of the median mass for this neutron star
would exclude most ``soft'' equations of state for dense neutron
matter. Millisecond pulsars (MSPs) appear to
have a much wider mass distribution than is found in double neutron
star systems; about half of these objects are significantly
more massive than $1.44\,M_{\sun}$. A possible cause is the much
longer episode of mass accretion necessary to recycle a MSP, which in
some cases corresponds to a much larger mass transfer.
\end{abstract}


\section{Introduction}\label{sec:intro}

Over the past 21 years, more than 130 pulsars have been discovered in
globular clusters (GCs)\footnote{For an updated list, see
  \url{http://www2.naic.edu/$\sim$pfreire/GCpsr.html}.}. Among the
first discoveries were PSR~B1516+02A and PSR~B1516+02B
\cite{wakp89}. Both are located in the GC NGC~5904. This cluster is
also known as M5 and we refer to these pulsars below as M5A and M5B.
M5A is an isolated millisecond pulsar (MSP) with a spin period of
5.55 ms. M5B is a 7.95-ms pulsar in a binary system with a
low-mass companion (see \S\ref{sec:masses}) and an orbital
period of 6.86 days. At the time of its discovery, this was the MSP
with the most eccentric orbit known ($e = 0.14$), this being $\sim\,
10^{4}\,-\,10^{5}$ times larger than that of MSP-White Dwarf
(WD) systems in the Galactic disk with similar orbital periods.

In the Galactic disk, 80\% of all known MSPs are found to be in binary
systems and, with the single exception of PSR~J1903+0327 (Champion et
al. 2008)\nocite{crl+08}, they are in low-eccentricity orbits with
WD companions. In GCs, gravitational interactions with neighboring
stars, or even exchange encounters, can produce binary systems with
eccentric orbits \cite{rh95}. These high eccentricities allow
the measurement of post-Keplerian parameters such as the rate of advance
of periastron ($\dot{\omega}$) or, in the future, the Einstein delay
($\gamma$) that are not normally measurable in MSP binaries in the
Galactic disk. If these effects are relativistic, they allow estimates of
the the total binary and component masses.

When Anderson~et~al.~(1997)\nocite{awkp97} published timing
solutions for M5A and B they used the eccentricity of M5B to detect
its periastron advance. However, the large relative uncertainty of the
measurement did not allow any astrophysically useful constraints on
the total mass of the binary. In this paper we report the results of
recent (2001 to 2008) 1.1-1.6~GHz (L-band) observations of M5. The
first 2001 observations were part of an Arecibo search for pulsars in
GCs, which found 11 new MSPs \cite{hrs+07}. Three of these
were found in M5, and subsequent observations of this GC were made
chiefly with the aim of timing these new discoveries (Stairs et
al. 2008, in preparation). M5A and B are in the same radio beam as
the new pulsars and they are clearly detectable in the L-band data,
permitting timing (see Figs. \ref{fig:profiles} and
\ref{fig:residuals}) of much better (M5A) or comparable (M5B) quality
to that obtained at 430 MHz by Anderson et al (1997). The whole dataset
now spans nearly 19 years and provides much improved timing parameters.

\section{Observations, data reduction and timing}
\label{sec:timing}

M5A and B were observed with the Arecibo 305-m radio telescope from the
time of their discovery in 1989 April until 1994 July using the 430-MHz
Carriage House line feed. For this, a 10-MHz band centered at 430~MHz
was used. The Arecibo correlation spectrometer made a 3-level
quantization of the signal and correlated this for a total of 128
lags. These data were then integrated for 506.58561 $\mu$s, and the
orthogonal polarizations added in quadrature before being written to
magnetic tape. The L-band  observations began in 2001 June, using the
``old'' Gregorian L-Wide receiver ($T_{sys}$ = 40~K at 1400 MHz). The 
``new'' L-Wide receiver ($T_{sys} = 25$K at 1400 MHz) has been used
since it was installed in the Gregorian dome in 2003 February. The
Wide-band Arecibo Pulsar Processors (WAPPs, Dowd, Sisk \& Hagen
2000\nocite{dsh00}) make a 3-level digitization of the voltages over a
100-MHz band for both (linear) polarizations, autocorrelating these
for a total of 256 lags. The data are then integrated for a total of
64$\,\mu$s and the orthogonal polarizations added in quadrature and
written to disk. At first, only one WAPP was available, and centered
the observing band at 1170~MHz or 1425~MHz. From 2003, three more
WAPPs have been available, and we now use three of them to
observe simultaneously at 1170, 1410 and 1510~MHz, the cleanest bands
within the wide frequency coverage of the new L-Wide receiver.

For all observations, the lags were Fourier transformed to generate
power spectra. For the L-band observations, the power spectra were
partially dedispersed at a dispersion measure (DM) of 29.5~cm$^{-3}$pc
and stored as a set of 16 sub-bands on the disks of the Borg computer
cluster at McGill University. At 1170~MHz, the partial
dedispersion introduces an extra smearing of 18 and 1.6$\mu$s for M5A
and B respectively. Adding these values in quadrature to the
dispersive smearing per channel (60.9 and 59.7 $\mu$s respectively),
we obtain a total dispersive smearing of 63.5 and 59.7 $\mu$s for M5A
and B respectively, i.e., the sub-banding introduces very little
extra smearing. The 430-MHz power spectra and L-band sub-bands were
dedispersed at the known DM of these pulsars and folded modulo their
spin periods. All the L-band data reported in this paper were
processed using the {\tt PRESTO} pulsar software
package\footnote{http://www.cv.nrao.edu/$\sim$sransom/presto}.

\begin{figure}
\epsscale{1.2}
\plotone{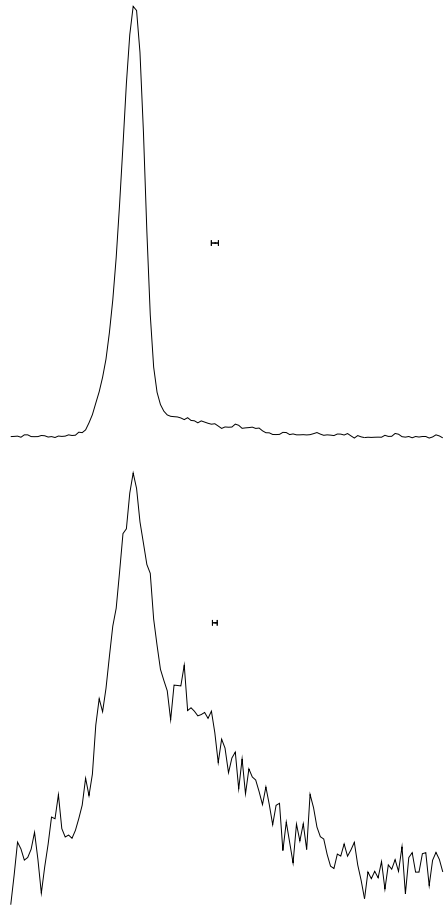}
\caption{\label{fig:profiles} Full-cycle pulse profiles for M5A (top)
  and M5B (bottom) at 1170 MHz, obtained by averaging the
  best detections of these pulsars at this frequency. The horizontal
  error bars denote the total time resolution of the system at
  1170~MHz: 90.2 and 87.5~$\mu$s for M5A and B respectively.}
\end{figure}

We added the best 1170 MHz detections of both pulsars to derive
``standard'' pulse profiles, and these are displayed in
Figure~\ref{fig:profiles}. A minimal set of Gaussian curves were
fitted to these profiles to derive synthetic templates for each
pulsar, and these were then cross-correlated
with each observation's pulse profile in the Fourier domain \cite{tay92}
to obtain topocentric times of arrival (TOAs). Adding all the
1170 MHz observations irrespective of their signal-to-noise ratio
(SNR, this varies from day to day because of diffractive interstellar
scintillation), we obtain a ``global'' pulse profile which has a lower SNR
than the ``standard'' pulse profile. We calculated the average flux
densities for both pulsars (see Table \ref{tab:parameters}) from the
off-pulse r.m.s. in their global profile, assuming a system equivalent
flux density of 3.5 Jy which is valid for 1170~ MHz at the high zenith
angles required to observe M5.

\begin{deluxetable*}{ l c c}
\tabletypesize{\scriptsize}
\tablecolumns{3}
\tablewidth{0pc}
\tablecaption{Parameters for two pulsars in NGC~5904}
\tablehead{
\colhead{Observation and flux parameters} &
\colhead{PSR~B1516+02A} &
\colhead{PSR~B1516+02B}}
\startdata
Start of 430-MHz observations \dotfill & \multicolumn{2}{c}{47635} \\
End of 430-MHz observations \dotfill & \multicolumn{2}{c}{49432} \\
Number of TOAs @ 430 MHz \dotfill & 86 & 82 \\
Residual rms @ 430 MHz  ($\mu$s) \dotfill & 49 & 114 \\
Start of L-band observations \dotfill & \multicolumn{2}{c}{52087} \\
End of L-band observations \dotfill & \multicolumn{2}{c}{54497} \\
Number of TOAs @ L-band \dotfill & 1278 & 162 \\
Uncertainty scale factor\tablenotemark{a} \dotfill & 1.50 & 1.05 \\
Residual rms @ L-band ($\mu$s) \dotfill & 9 & 72 \\
Average flux density @ 1170 MHz (mJy) \dotfill & 0.155 & 0.027 \\
\hline
\multicolumn{1}{c}{Ephemeris} & \multicolumn{2}{c}{} \\
\hline
Reference Epoch (MJD) \dotfill & 54000 & 54000 \\
Right Ascension, $\alpha$ (J2000) \dotfill &
$15^{\rm h}18^{\rm m}33\fs 32307(6)$\tablenotemark{b} &
$15^{\rm h}18^{\rm m}31\fs 4625(8)$ \\
Declination, $\delta$ (J2000)\dotfill  &
$+02^\circ 05\arcmin 27\farcs 435(3)$ &
$+02^\circ 05\arcmin 15\farcs 30(3)$ \\
Proper motion in $\alpha$, $\mu_{\alpha}$ (mas yr$^{-1}$, J2000)
\dotfill & 4.6(4) & 3.4(1.2) \\
Proper motion in $\delta$, $\mu_{\delta}$ (mas yr$^{-1}$, J2000)
\dotfill & $-$8.9(1.0) & $-$11.8(2.8) \\
Spin frequency, $\nu$ (Hz) \dotfill &
180.063624055103(3) &
125.83458757935(6) \\
Time derivative of $\nu$, $\dot{\nu}$ (10$^{-15}$ Hz s$^{-1}$) \dotfill &
$-$1.33874(4) &
0.05233(15) \\
Dispersion Measure, DM (pc\,cm$^{-3}$) \dotfill & 30.0545(10) & 29.46(3) \\
Orbital period, $P_b$ (days) \dotfill & \nodata & 6.8584538(3) \\
Projected size or orbit, $x$ (l-s) \dotfill & \nodata & 3.04857(2) \\
Orbital eccentricity, $e$ \dotfill & \nodata & 0.137845(10) \\
Time of passage through periastron, $T_0$ (MJD) \dotfill & \nodata & 54004.02042(15) \\

Longitude of periastron, $\omega$ ($^\circ$) \dotfill & \nodata & 359.898(8) \\
Rate of advance of periastron, $\dot{\omega}$ ($^\circ$ yr$^{-1}$)
\dotfill & \nodata & 0.0142(7) \\
Second time derivative of $\nu$, $\ddot{\nu}$ ($10^{-27}$ Hz s$^{-2}$) \dotfill &
[$-0.6 \pm 0.5$]\tablenotemark{c} & [$4.4 \pm 5.4$] \\
Time derivative of $x$, $\dot{x}$ ($10^{-12}$ l-s/s) \dotfill &
\nodata & [$-\,0.12\,\pm\,0.09$] \\
Time derivative of $P_B$, $\dot{P_B}$ ($10^{-12}$) \dotfill &
\nodata &  [$15\,\pm\,31$] \\
\hline
\multicolumn{1}{c}{Derived parameters} & \multicolumn{2}{c}{}\\
\hline
Spin period, $P$ (ms) \dotfill & 5.55359254401089(11) & 7.946940656275(4) \\
Time derivative of $P$, $\dot{P}$ (10$^{-21}$ s s$^{-1}$) \dotfill &
41.2899(13) & $-$3.306(10) \\
Mass function, $f\,(M_{\sun})$ \dotfill & \nodata & 0.000646723(13) \\
Total system mass, $M$ $(M_{\sun})$ \dotfill & \nodata & 2.29(17) \\
Maximum pulsar mass, $M_{p, \rm max}\,(M_{\sun})$ \dotfill & \nodata & 2.52 \\
Minimum companion mass, $M_{c, \rm min}\,(M_{\sun})$ \dotfill & \nodata & 0.13 \\
\enddata
\tablenotetext{a}{This is the scale factor for TOA uncertainties
  required to achieve a reduced $\chi^2$ of unity in the solution.}
\tablenotetext{b}{1-$\sigma$ uncertainties are presented in
  parenthesis. These are twice the estimate obtained using the
  Monte-Carlo bootstrap method.}
\tablenotetext{c}{Values in square brackets are not
  considered to be significant. They were not fit when
  determining the remaining timing parameters.
}
\label{tab:parameters}
\end{deluxetable*}

\begin{figure*}
\epsscale{0.7}
\plotone{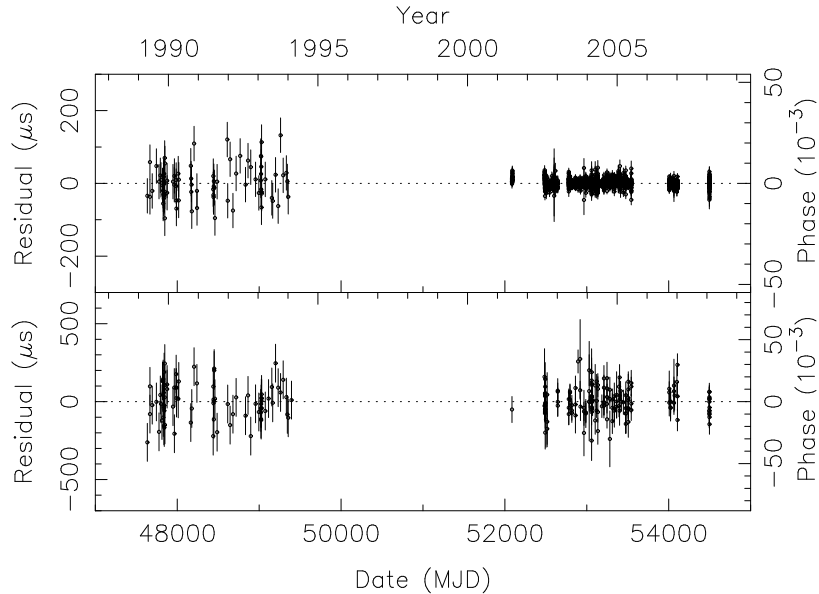}
\caption{\label{fig:residuals} Timing residuals for M5A (upper) and
  M5B (lower), obtained with the ephemerides in
  Table~\ref{tab:parameters}. The large gap in time coverage is mostly
  due to the Arecibo upgrade.}
\end{figure*}

The TOAs were analyzed with
{\tt TEMPO}\footnote{\url{http://www.atnf.csiro.au/research/pulsar/tempo/}},
using the DE~405 Solar System ephemeris \cite{sta98b} to model the
motion of the Arecibo radio telescope relative to the Solar System
Barycenter. The orbital parameters for M5B were modeled using the
Damour \& Deruelle orbital model \cite{dd85,dd86}. For most of the
early 430-MHz TOAs we have no reliable uncertainty
estimates. Therefore, in order to achieve a reduced $\chi^2$ of 1 for
both pulsars, we attributed a constant uncertainty to the 430-MHz TOAs
that is similar to their unweighted rms. These times are listed in
Table \ref{tab:parameters}. With these TOAs, we obtain
timing parameters that are virtually identical to those obtained by the
previous analysis (Anderson et al. 1997). For the new L-band data, we
find the TOA uncertainties to be under-estimated in the case of
M5A. Multiplying these by a factor of about 1.5, we achieve a reduced
$\chi^2$ of 1 for its L-band TOAs. In the case of M5B, this factor is
1.05. These values are similar to those derived for other MSPs
timed with the same software \cite{frb+08}.

The resulting timing parameters and their 1-$\sigma$ uncertainties are
presented in Table \ref{tab:parameters}. We estimate these
uncertainties to be twice the 1-$\sigma$ Monte-Carlo bootstrap
\cite{et93,ptvf92} uncertainties.  We discuss the validity of this
choice for the particular case of the periastron advance of
M5B in \S\ref{sec:omega-dot}. All the parameters that vary in time
($\alpha$, $\delta$, $\nu$ and $\omega$) are estimated for the
arbitrary epoch, MJD = 54000 (2006 September 22); $T_0$ was the first
periastron passage to occur after that date. The post-fit timing
residuals are essentially featureless at the present timing accuracy
(see Fig. \ref{fig:residuals}). The large gap without measurements
between the early 430-MHz data and later L-band data is in
part due to the Arecibo upgrade of the late 1990's. We included an
arbitrary time step between these two datasets in the fit. We have tested
the timing solution by introducing extra pulsar rotations between the
two datasets, but these are always absorbed by this arbitrary time
step, with no other changes in the fitted timing parameters.

The positions, periods and period derivatives that we have obtained are
consistent with those of Anderson et al (1997). In what follows, we
analyze solely the newly measured parameters: the proper motions and
the rate of advance of periastron of M5B.

\section{Proper motions}

In the reference frame of a GC, the rms of the velocities of its
pulsars along the orthogonal axes perpendicular to the line of sight
should be the same as the rms of their velocities along the line of
sight. The stellar rms velocity along the line of sight at the center
of M5 is 7.15~km~s$^{-1}$ \cite{web85}; the rms of the pulsar
velocities should be smaller given the larger masses of the neutron
stars (NSs). At the distance of M5, 7.5~kpc \cite{har96}, the stellar rms
velocity represents a proper motion rms of only
0.2~mas~yr$^{-1}$. Given the present measurement precision, the
estimated pulsar proper motions should be
mutually consistent and reflect only the proper motion of the GC.
The proper motion measurements in Table~\ref{tab:parameters} are
indeed $\sim 1\,\sigma$ consistent with each other; from their weighted average
we derive a proper motion for M5 of $\mu_{\alpha}\,=\,(4.3\,\pm\,0.4)$
mas yr$^{-1}$ and $\mu_{\delta}\,=\, (-9.6\,\pm\,1.0)$ mas yr$^{-1}$.

M5 is one of the four GCs in the Galaxy for which optical proper
motion measurements have not provided consistent (i.e., agreeing within
the formal uncertainty estimates) results
(Dinescu, Girard \& van Altena 1999)\nocite{dga99}. Our M5
proper motion measurement is in marginal agreement with the values derived by
Scholz et al. [1996, $\mu_{\alpha}\,=\,(6.7\,\pm\,0.5)$ mas yr$^{-1}$
and $\mu_{\delta}\,=\, (-7.8\,\pm\,0.4)$ mas
yr$^{-1}$]\nocite{soh+96}, but is in good
agreement with the values derived from Hipparcos [Odenkirchen et
al. 1997, $\mu_{\alpha}\,=\,(3.3\,\pm\,1.0)$ mas yr$^{-1}$ and
$\mu_{\delta}\,=\, (-10.1\,\pm\,1.0)$ mas yr$^{-1}$]\nocite{obgt97}.

\begin{figure}
\epsscale{1.15}
\plotone{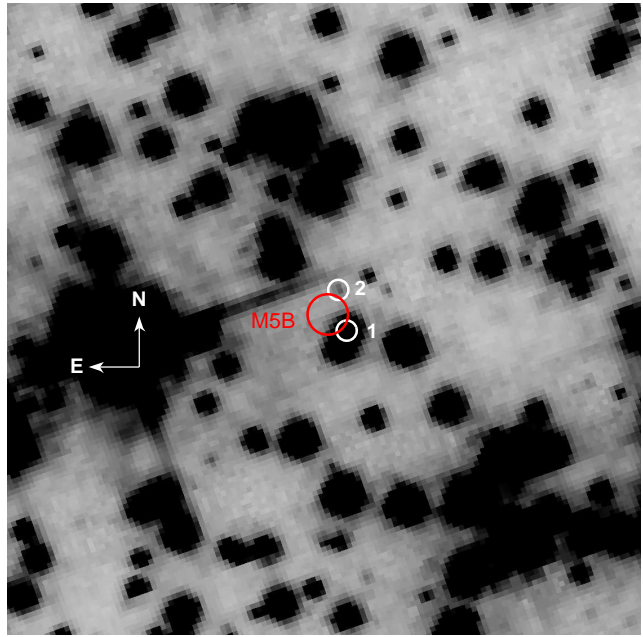}
\caption{\label{fig:optical_M5B} A small portion of a GO-10120
{\em HST} image of M5, centered on the position of M5B, in
negative. The picture was taken through the F435W filter of the Advanced
Camera for Surveys/Wide Field Camera. The dark circle indicates the
position of M5B and the radius (0.2\arcsec) corresponds to the
2$\sigma$ uncertainty in the absolute astrometry of the image. The
nearest stars (white circles) lie at 2.7 and 3.0$\sigma$ from the
radio position; these are indicated as A and B and are discussed in \S
\ref{sec:optical}.}
\end{figure}

\section{Search for the optical counterpart of M5B}
\label{sec:optical}

We have used the astrometric information on M5B to search for an
optical counterpart in archival {\em HST} ACS/WFC data from programs GO
10120 and GO 10615. The GO-10120 images were taken on 2004 August 1
through the F435W, F625W and F658N filters. Since the uncertainty in
the absolute astrometry of {\em HST} data is 1--2\arcsec, we first tie
the astrometry of the GO-10120 ACS images to the ICRS frame using
UCAC2 stars \cite[positional accuracy $\lesssim$0.070\arcsec~down to
the magnitude limit of the UCAC2 catalog,][]{zachea04}. Since UCAC2
standards in the small ACS field (3.4\arcmin$\times$3.4\arcmin) are
scarce, we use ground-based imaging of M5 to
derive secondary standards. We retrieved from the public archive of
the 2.5 m Isaac Newton Telecope (INT) a 30-s Sloan-r Wide
Field Camera image taken on 2004 June 8 and processed only the chip
that contained the core of M5 (field of view
$\sim$23\arcmin$\times$11\arcmin). Astrometric calibration of this
image was achieved using 308 UCAC2 stars with positions corrected for
proper motion to the epoch of the INT image. After fitting for shift,
rotation angle, scale factor and distortions, the final solution has
rms residuals of 0.050\arcsec~in right ascension and 0.047\arcsec~in
declination. We selected a set of 198 secondary standards from
unsaturated and relatively isolated stars in the INT image. These were
used in turn to compute an astrometric solution for the short (70s)
F435W distortion-corrected (using the {\em multidrizzle} software)
exposure. The resulting fit for shift, rotation angle and scale factor
has rms residuals of 0.017\arcsec~in right ascension and
0.014\arcsec~in declination. We estimate the final 1$\sigma$ accuracy
of our ACS absolute astrometry as the quadratic sum of the errors in
the UCAC2 astrometry, the UCAC2--INT tie and the INT--{\em HST} tie,
i.e.~0.1\arcsec~(or 2 ACS pixels).

The position of M5B at the epoch of the GO-10120 observations is shown
in Fig.~\ref{fig:optical_M5B}. No optical sources are detected within
the 2-$\sigma$ error circle of the radio position (this only includes
the uncertainty in the absolute astrometry as the uncertainties in the
radio position are negligible). The photometry shows the nearest
sources---indicated as ``1'' and ``2'' at distances 2.7 and 3.0\,$\sigma$,
respectively---to be
main-sequence (MS) stars located $\sim$1.3 and $\sim$6.1 mag (in F435W)
below the MS turnoff. Using the M5 turnoff magnitude from
Sandquist et al.~(1996) \nocite{sandea96} and assuming $B \approx
F435W$, the 14-Gyr isochrones from Bergbusch \& Vandenberg (1992)
\nocite{bergvand92} imply that ``1'' and ``2'' are $\sim$0.75 and
$\sim$0.4$M_{\sun}$ stars, respectively (for a distance modulus
$(m-M)=14.41$ and $E(B-V)=0.03$, see Sandquist et al 1996). To check
for fainter stars, we stacked F435W images taken with fraction-pixel offsets
from {\em HST} program GO 10615 (2006 Feb 15)
into a deep (8500 s) high-resolution (twice-oversampled)
masterframe. Psf photometry revealed no additional stars within 3$\sigma$
of M5B. A conservative upper limit for the detection limit is
m$_{F435W}\approx 26-26.5$ mag which corresponds to
$\sim$0.25--0.3$M_{\sun}$ for MS stars in M5 (although close to
star ``1'' the sensitivity is lower).

In summary, we cannot exclude stars ``1'' or ``2'' as counterparts of
M5B. However, the astrometry suggests that it is more likely that the
true optical counterpart is fainter than our detection limit. That
implies that the companion is either a WD or a faint, low-mass MS star.

\section{Periastron advance of M5B}
\label{sec:omega-dot}

We have determined a highly significant estimate for the rate of
advance of periastron of M5B: $\dot{\omega}\,=\,0.0142(7)^\circ \rm
yr^{-1}$. The 1-$\sigma$ estimate provided directly by {\tt TEMPO} is
$\dot{\omega}\,=\,0.01422(43)^\circ \rm yr^{-1}$. To verify that these
values are realistic, we have kept $\dot{\omega}$ fixed, and fitted all
the remaining timing parameters, recording the resulting
$\chi^2$. Doing this for a range of values of $\dot{\omega}$, we
obtain the 1-$\sigma$ uncertainty as the half-width of the region
where $[\chi^2 (\dot{\omega}) - \chi^2 (\dot{\omega}_{\rm min}) < 1]$,
$\dot{\omega}_{\rm min}$ being the value that minimizes $\chi^2$
\cite{sna+02}. The result is $\dot{\omega}\,=\,0.001422(43)^\circ \rm
yr^{-1}$. Estimating all parameters using a Monte-Carlo Bootstrap
algorithm (\S \ref{sec:timing}) we obtain
$\dot{\omega}\,=\,0.01422(36)^\circ \rm yr^{-1}$. Furthermore, the
lack of significant higher derivatives of the spin frequency (see
Table~\ref{tab:parameters})  suggests that, with the present timing
precision, we are unable to detect any of this pulsar's timing noise
which can contaminate the physical interpretation of its timing
uncertainties. This is a necessary pre-condition for an accurate
estimation of parameter uncertainties. We therefore believe that the
$1 \sigma$ {\tt TEMPO} uncertainty estimates are essentially
accurate. We choose, however, to be more conservative by making our
1-$\sigma$ uncertainties twice as large as the values suggested by the
Monte-Carlo method (see Table~\ref{tab:parameters}). This caution is
due to our use of two different datasets with a large time gap between
them.

This time gap is common to many Arecibo timing data sets, such as that
of PSR~J0751+1807. Nice et al. (2005)\nocite{nss+05} claimed a mass of
$2.1\,M_{\sun}$ for that pulsar. This claim has recently been
retracted, the latest estimate for the pulsar mass now being
$1.26^{+0.14}_{-0.12} M_{\sun}$ (1$\sigma$, Nice 2007\footnote{See
  also http://www.ns2007.org/talks/nice.pdf}).
The problem with the earlier estimate was not related to the gap as the
new estimate is based on essentially the same data. Its cause
was the use of a necessarily imperfect ephemeris when folding
the very earliest data. This resulted in orbital-dependent smearing of
the pulse profiles that led to an error in the
calculation of the orbital phase for the earliest data and an
over-estimate of the orbital period decay. Because those data were
folded online, this problem could only be solved by ignoring the
earliest TOAs. Our 430-MHz data contained the signals of more than one
pulsar, so we had to record them to tape. This allowed us to re-fold them
iteratively after the timing solution had been obtained (Anderson et
al. 1997). After updating the timing solution to 2008, we have also
re-folded all our L-band data; none of the pulse profiles used in this
work are therefore smeared due to imprecise folding.

\subsection{Is $\dot{\omega}$ relativistic?}
\label{sec:relativistic}

The possible contributions to $\dot{\omega}$ in a system containing a pulsar
and an extended star have been studied in detail by Lai, Bildsten \& Kaspi
(1995)\nocite{lbk95}. In their analysis of the binary pulsar
PSR~J0045$-$7319 they concluded that the only likely contribution to
$\dot{\omega}$ in such systems is from rotational deformation of the
companion. If we assume that PSR~J0045$-$7319 has a mass of
$1.4\,M_{\sun}$, the companion mass is $8.8\,M_{\sun}$ and its radius
$R_c \sim 6.4\,R_{\sun}$ \cite{bbs+95}; the orbital separation $a$ is
$\sim 126\,R_{\sun}$. Using these values they reached the conclusion
that the contribution to $\dot{\omega}$ from tidal deformations, which
are proportional to $(R_c / a)^3 \simeq 1.32 \times 10^{-4}$, is not
significant in that system. For M5B, again assuming a pulsar mass of
$1.4\,M_{\sun}$ with a MS companion of maximum mass $0.3\,M_{\sun}$
(see \S~\ref{sec:optical}) and a radius $R_c = 0.3\,r_{0.3}\,R_{\sun}$
(where $r_{0.3} \sim 1$), then the orbital inclination $i$ is $\sim
24^{\circ}$, $a \sim 18\,R_{\sun}$ and $(R_c / a)^3 \simeq 4.5 \times
10^{-6} r_{0.3}^{3}$. This is 30 times smaller than for
PSR~J0045$-$7319, becoming even less significant for smaller MS
companion masses and correspondingly higher inclinations.

Using equations 68 and 79 of Wex (1998) we can estimate the
contribution to $\dot{\omega}$ due to rotational deformation:
\begin{equation}
\label{eq:rot}
\dot{\omega}_{\rm rot} = n \frac{k R_c^2 \hat{\Omega}^2}{a^2 (1 - e^2)^2}
\left( 1 - \frac{3}{2} \sin^2 \theta + \cot i \sin \theta \cos \theta
\cos \Phi_0 \right),
\end{equation}
where $n$ is the orbital angular frequency ($2 \pi / P_b = 1.06 \times
10^{-5} \rm rad\,s^{-1}$), $k$ is the gyration radius (for a homogeneous
sphere this is 0.63, while for any centrally condensed objects 
this will always be smaller: it is about 0.2 for a completely
convective star), $\hat{\Omega}$ is the rotation rate relative to break-up,
$\theta$ is the angle between the rotational and orbital angular momenta
(if the companion is non-degenerate, these tend to be aligned, so $\theta = 0$)
and $\Phi_0$ is the longitude of the ascending node in a reference frame
defined by the total angular momentum vector (see Fig.~9 of Wex 1998).

For the situation discussed above (a 1.4-$M_{\sun}$ pulsar with the
largest possible MS companion, $R_c = 0.3 r_{0.3} R_{\sun}$), we have
$R_c / a = 16.55 \times 10^{-3} r_{0.3}$. We can also calculate the
break-up angular velocity:
$\Omega_{\rm max} = \sqrt{GM_c/R_c^3} = 1/ 0.3 \sqrt{GM_{\sun} /
  (r_{0.3} R_{\sun})^3} = 2.09 \times 10^{-3} r_{0.3}^{(-3/2)}\,
\rm rad\,s^{-1}$. If the companion's rotation is tidally locked to
its orbit around the pulsar, then
$\hat{\Omega} = n / \Omega_{\rm max} = 5.06 \times 10^{-3} r_{0.3}^{(3/2)}$.
Therefore, $\dot{\omega}_{\rm rot} = 2.65 \times 10^{-15}
r_{0.3}^{5} {\rm rad\,s^{-1}} = 4.79 \times 10^{-6} (r_{0.3}^{5})^{\circ}
\rm yr^{-1}$ (or three times this if the companion were to be a
homogeneous sphere). This is $\sim 3 \times 10^{3}$ times smaller than
the value we dedetermined (\S \ref{sec:omega-dot}). If the companion
were to be significantly distended for its mass (i.e., if $r_{0.3} >
1$), as is the case for the companion of PSR~J1740$-$5340
\cite{fpds01}, then $\dot{\omega}_{\rm rot}$ could be
significant. This scenario can be excluded, since no optical
counterpart is readily detectable within $\sim 2.5 \sigma$ of the pulsar
(see \S \ref{sec:optical}).

If the companion were to be a WD, then it could be more massive than
$0.3\,M_{\sun}$ and still evade optical detection. Irrespective of its mass,
the contribution to $\dot{\omega}$ from the tidal deformation of a WD
is negligible, but that is not necessarily the case for the contribution
from rotational deformation. As an example, we re-calculate
$\dot{\omega}_{\rm rot}$ for a 0.3-$M_{\sun}$ WD. For WDs, we have $k = 0.45$
\cite{lp98}, more than twice as large as for fully convective
stars. For WDs $r_{0.3}$ is of the order of 0.1, i.e., the $(R_c/a)^2$
term in eq.~\ref{eq:rot} would be $\sim 10^{2}$ times smaller than
discussed for $r_{0.3} \sim 1$. However, a WD companion is not
likely to be tidally locked. If it were spinning fast, then
$\hat{\Omega} \sim 1$. While there is no special {\it a priori} reason
why this should be true, it is a possibility that cannot be excluded. This
would mean that $\hat{\Omega}^2$ could be $\sim 4 \times 10^{4}$ times
larger than discussed above, with $\dot{\omega}_{\rm rot}$ similar to the 
observed $\dot{\omega}$. This is particularly so for the larger-sized WDs
(those with the lowest masses). Following Splaver et al. (2002), we
note first that if the companion is not tidally locked, the angular
momenta of the orbit and companion spin will probably not be aligned
($\theta \neq 0$). In this case, the spin of the companion will induce
a precession of the orbital plane. This will cause a change in $i$,
affecting the projected semi-major axis of the orbit which will vary
with a rate $\dot{x}$. Rewriting  equation 81 of Wex
(1998)\nocite{wex98}, we can relate $\dot{\omega}_{\rm{rot}}$ to $\dot{x}$:
\begin{equation}
\dot{\omega}_{\rm rot} = 
  \frac{\dot{x}}{x}\,
  \left(
  \tan i
    \frac{1-\frac{3}{2}\sin^2\theta}{%
          \sin\theta\cos\theta\sin\Phi_0}
      + \cot \Phi_0
  \right).
\end{equation}
This equation has the advantage that it does not depend on the mass
(or the nature) of the companion.
Thus, our observed 2-$\sigma$ upper limit of  $|\dot{x}/x|< 9.6 \times
10^{-14}\,\rm s^{-1}$ implies $|\dot{\omega}_{\rm{rot}}|\,<\,(1.7 \,
\times\,10^{-4})^\circ$\,yr$^{-1}$ times a geometric factor. In 80\% of
cases this geometric factor will be smaller than 10 and
the upper limit for $\dot{\omega}_{\rm{rot}}$ is similar to the present
measurement uncertainty for $\dot{\omega}$.

To summarize, $\dot{\omega}_{\rm rot}$ can only be significant
if the companion is degenerate, rotating near breakup velocity, and
with its rotational angular momentum nearly aligned with the orbital
angular momentum, making $\dot{x}$ undetectable. Otherwise
$\dot{\omega}$ is relativistic.

\begin{figure*}
\plotone{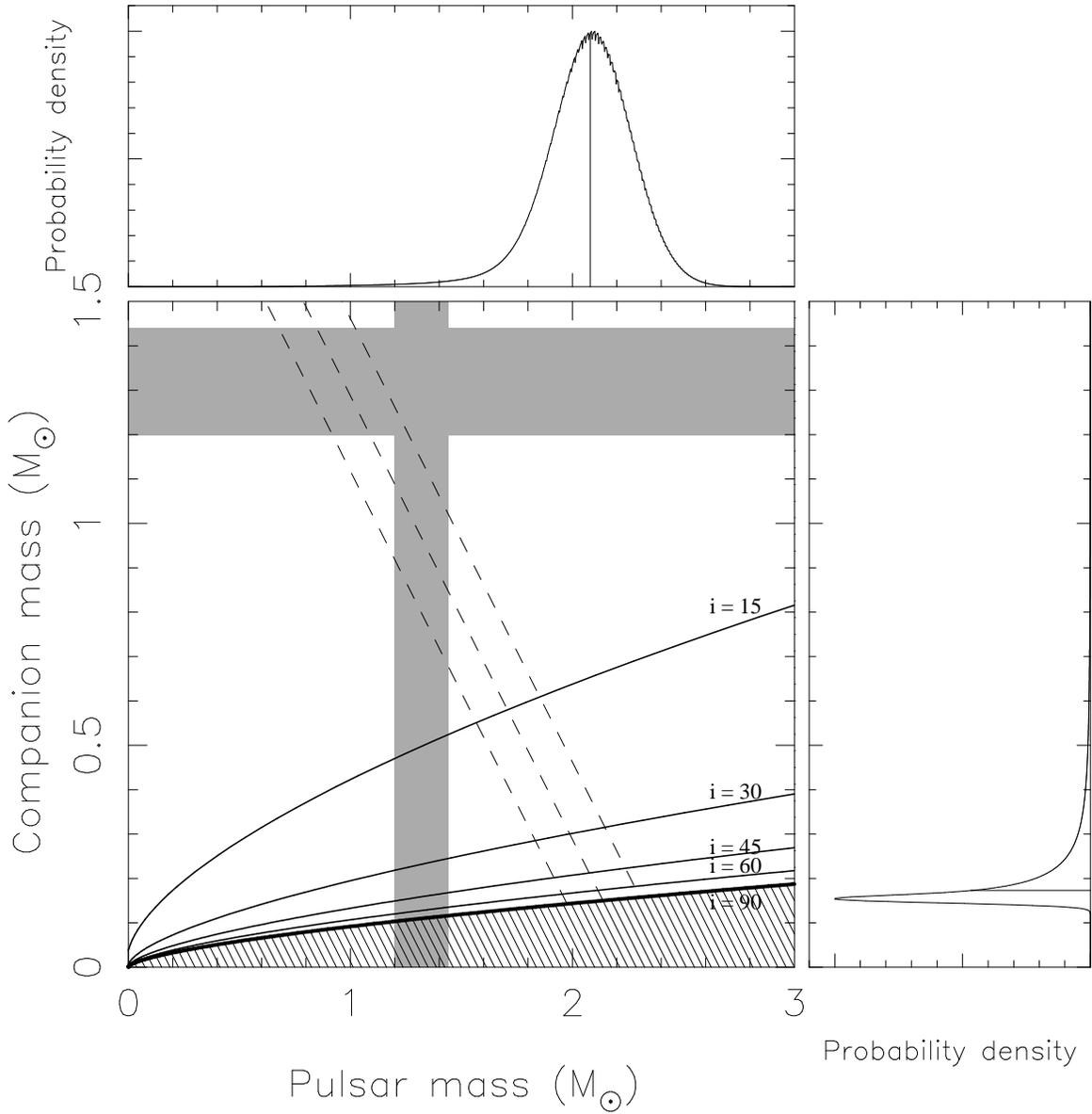}
\caption{\label{fig:mass_mass} Constraints on the masses of M5B
  and its companion. The hatched region is excluded by knowledge of
  the mass function and by $\sin i \leq 1$.  The diagonal dashed lines
  correspond to a total system mass that causes a general-relativistic
  $\dot{\omega}$ equal to, or within 1-$\sigma$ of, the measured value.
  The six solid curves indicate constant inclinations. We also
  display the probability density function for
  the mass of the pulsar ({\em top}) and the mass of the companion
  ({\em right}), and mark the respective medians with vertical
  (horizontal) lines. For comparison, the gray bars indicate the range
  of ``normal'' NS masses (see \S\ref{sec:masses}).}
\end{figure*}

\subsection{Binary, pulsar and companion masses}
\label{sec:masses}

When $\dot{\omega}$ is solely due to the effects of general
relativity, we can measure the total mass of a binary system:
\begin{equation}
\label{eq:total_mass}
{M} = \left(\frac{P_b}{2 \pi}\right)^{5/2}
\left[\frac{(1 - e^2)\,\dot{\omega}}{3}\right]^{3/2}
\left(\frac{1}{T_\sun}\right),
\label{eq:totmass}
\end{equation}
where $T_\sun \equiv G M_{\sun} / c^3\,=\,4.925490947 \mu$s. For M5B,
we obtain $M\,=\,(2.29\,\pm\,0.17)\,M_{\sun}$. For the nominal value
of $\dot{\omega}$ and a median $i$ of 60$^\circ$,
the mass of the companion is 0.173\,$M_{\sun}$ and the mass of the
pulsar is 2.11\,$M_{\sun}$. This is well above all NS masses that have
been precisely measured to date.

We calculated a 2-D probability distribution function (pdf) for the
mass of the pulsar and the mass of the companion, assuming that the
pdf for $\dot{\omega}$ is a Gaussian with the half-width equal to
the 1-$\sigma$ uncertainty listed in Table~\ref{tab:parameters} and an
{\em a priori} constant probability for $\cos i$. The two-dimensional
pdf is then projected in both dimensions, resulting in 1-D pdfs for
the mass of the pulsar and the mass of the companion. These are
displayed graphically in Fig.~\ref{fig:mass_mass}. The pulsar
definitely has a mass smaller than 2.52\,$M_{\sun}$, and the
companion has a mass larger than 0.13\,$M_{\sun}$, the median and
1-$\sigma$ limits for the pulsar and companion mass are
$2.08^{+0.18}_{-0.19} M_{\sun}$ and $0.172^{+0.107}_{-0.023} M_{\sun}$
respectively. There is a 99\%, 95\% and 90\% probability that the
pulsar is more massive than 1.38, 1.72 and 1.82\,$M_{\sun}$
respectively. There is a 0.77\% probability that $i$ is low
enough to make the NS mass fall within the range of the components of
double neutron star (DNS) systems: from 1.20~$M_{\sun}$ measured for
the companion of PSR~J1756$-$2251 \cite{fkl+05} to 1.44~$M_{\sun}$
measured for PSR~B1913+16 \cite{wt03}. For M5B, assuming its nominal
value of $\dot{\omega}$, these mass limits would imply that
$7.9^\circ\,<\,i\,<\,10.2^{\circ}$.

\begin{deluxetable*}{ l c r c c c c c c c c}
\tabletypesize{\footnotesize}
\tablecolumns{10}
\tablewidth{0pc}
\tablecaption{Millisecond Pulsar Masses}
\tablehead{
\colhead{Name PSR} &
\colhead{GC} &
\colhead{$P$ (ms)} &
\colhead{$P_b$ (days)} &
\colhead{$e$} &
\colhead{$f/M_{\sun}$} &
\colhead{$M/M_{\sun}$\tablenotemark{a}} &
\colhead{$M_c/M_{\sun}$} &
\colhead{$M_p/M_{\sun}$} &
\colhead{Method\tablenotemark{b}} &
\colhead{Ref.\tablenotemark{c}}}
\startdata

\multicolumn{11}{c}{Selected MSP Mass Measurements} \\

\hline

J0751+1807 & - & 3.47877 & 0.26314 & 0.00000 & 0.0009674 & - & - &
$1.26^{+14}_{-12}$ & $\dot{P_b}$, $s$ & 1 \\

J1911$-$5958A & NGC~6752 & 3.26619 & 0.83711 & $<$0.00001 &
0.002688 & 1.58$^{+0.16}_{-0.10}$ & 0.18(2) & $1.40^{+0.16}_{-0.10}$ & Opt. &
2 \\

J1909$-$3744 & - & 2.94711 & 1.53345 & 0.00000 & 0.003122 & $1.67^{+3}_{-2}$ & 0.2038(22) & $1.47^{+3}_{-2}$ & $r,s$ & 3 \\

J0437$-$4715 & - & 5.75745 & 5.74105 & 0.00002 & 0.001243 & 2.01(20) &
0.254(14) & 1.76(20) & $r,s$ & 4 \\

J1903+0327 & - & 2.14991 & 95.1741 & 0.43668 & 0.139607 & 2.88(9) & 1.07(2) & 1.81(9) & $\dot{\omega}$, $s$ & 5 \\

\hline

\multicolumn{11}{c}{Binary systems with indeterminate orbital inclinations} \\

\hline

J0024$-$7204H & 47~Tucanae & 3.21034 & 2.35770 & 0.07056 & 0.001927 & 1.61(4) &
 $> 0.164$ & $< 1.52$ & $\dot{\omega}$ & 6 \\

J1824$-$2452C & M28 & 4.15828 & 8.07781 & 0.84704 & 0.006553 & 1.616(7) &
 $> 0.260$ & $< 1.367$ & $\dot{\omega}$ & 7 \\

\hline

J1748$-$2446I & Terzan~5 & 9.57019 & 1.328 & 0.428 & 0.003658 & 2.17(2) &
$> 0.24$ & $< 1.96$  & $\dot{\omega}$ & 8 \\

J1748$-$2446J\tablenotemark{d} & Terzan~5 & 80.3379 & 1.102 & 0.350 & 0.013066 & 2.20(4) &
$> 0.38$ & $< 1.96$  & $\dot{\omega}$ & 8 \\

B1516+02B & M5 & 7.94694 & 6.85845 & 0.13784 & 0.000647 & 2.29(17) &
$> 0.13$ & $< 2.52$  & $\dot{\omega}$ & \S\ref{sec:masses} \\

J0514$-$4002A\tablenotemark{e} & NGC~1851 & 4.99058 & 18.7852 &
0.88798 & 0.145495 & 2.453(14) & 
$> 0.96$ & $< 1.52$ & $\dot{\omega}$ & 9 \\

J1748$-$2021B & NGC~6440 & 16.76013 & 20.5500 & 0.57016 & 0.000227 & 2.91(25) &
$> 0.11$ & $< 3.3$ & $\dot{\omega}$ & 10 \\

\enddata
\tablenotetext{a}{Binary systems are sorted according to the total
  estimated mass $M$.}
\tablenotetext{b}{Methods are: $\dot{P_b}$ - relativistic orbital
  decay, $r,s$ - Shapiro delay ``shape'' and ``range'', ``Opt'' -
  optically derived mass ratio, plus mass estimate based on spectrum
  of companion, $\dot{\omega}$ - precession of periastron.}
\tablenotetext{c}{References are 1: \cite{nice07} (total and companion
  masses not provided), 2: \cite{bkkv06},
  3: \cite{hbo06}, 4: \cite{vbs+08}, 5: \cite{crl+08}, 6:
  \cite{fck+03}, 7: \cite{brf+08}, 8: \cite{rhs+05}, 9: (Freire, Ransom
  \& Gupta 2007)\nocite{frg07}; 10: \cite{frb+08}}
\tablenotetext{d}{This pulsar is not technically a MSP, its spin
  period is longer than those found in most DNS systems. However,
  given the similarity of its orbital parameters to those of
  Terzan~5~I, we assume that it had a similar formation history.}
\tablenotetext{e}{Because of its large companion mass and eccentricity,
this system is thought to have formed in an exchange interaction
  \cite{fgri04}.}
\label{tab:msp_masses}
\end{deluxetable*}

\section{Statistical evaluation of mass measurements}

M5B has the second largest mass estimate among all known NSs
after PSR~J1748$-$2021B (NGC 6440B). Because of indeterminate
orbital inclinations, all mass estimates based solely on $\dot{\omega}$
are probabilistic statements: one more PK parameter is necessary
to have an unambiguous determination of $i$ and $M_p$. No such parameters
have yet been measured for M5B, NGC~6440B or any other eccentric MSP
binaries in GCs; this is in some cases due to their low timing
precision (like M5B, which is faint and has a broad pulse profile),
and in others to their small timing baselines (like
NGC~6440B). Nevertheless, unambiguous upper limits for the pulsar
masses and lower limits for
the companion masses can always be obtained from a measurement
of a relativistic $\dot{\omega}$ alone (see Table~\ref{tab:msp_masses}).
Furthermore, in systems where the mass function is very small and the
total binary mass is very large (as for M5B and NGC~6440B) there is a
much greater probability that most of the mass of the binary belongs to
the pulsar itself, as described in \S\ref{sec:masses}.

\begin{figure}
\epsscale{1.15}
\plotone{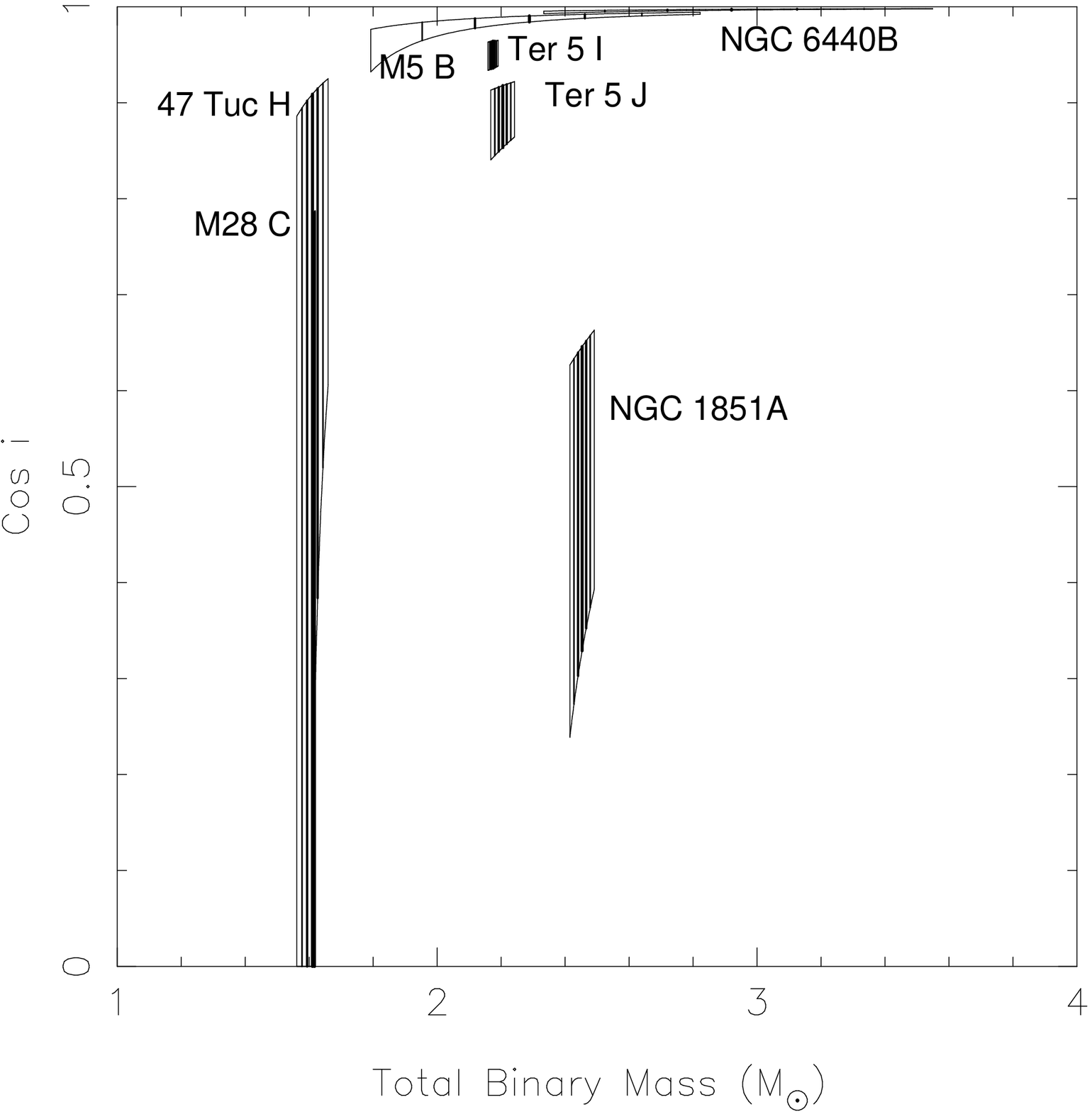}
\caption{\label{fig:binaries}
Cosine of the orbital inclination $i$ as a function of total binary
mass for the eccentric MSP binaries in GCs. For
each binary, the upper curve assumes a pulsar mass of $1.2\,M_{\sun}$
and the lower curve $1.44\,M_{\sun}$. These are different for each
binary because of their different mass functions. The vertical lines
indicate the total masses corresponding to $\dot{\omega}$ and its
$\pm\,1, 2$ and $3-\sigma$ uncertainties, based on the most recent
timing. If we assume randomly oriented orbits, then for any given total
mass value the vertical distance between the two curves gives us the
probability of the pulsar mass falling within the $1.2 - 1.44\,M_{\sun}$
range. With the exception of NGC~1851A, these probabilities are
significantly smaller for the more massive systems.}
\end{figure}

\subsection{Evidence for high average neutron star masses}

An interesting feature of the eccentric binary MSPs in GCs is
that as the binary mass increases, the mass function $f$ does not
increase (see Table~\ref{tab:msp_masses}). The exception is
NGC~1851A; a system thought to have resulted from an exchange
interaction \cite{fgri04}. If these binaries were to have
$M_p\,<\,1.44\,M_{\sun}$ then the increase in total mass would be due
to higher companion masses, resulting in a general trend to higher
mass functions. This is generally not the case.

If we assume that all these GC MSPs have ``normal'' masses (between
1.2 and $1.44\,M_{\sun}$), we can then calculate the orbital
inclinations of these binaries from their total masses and mass
functions. These are displayed graphically in
Fig.~\ref{fig:binaries}. Of the five massive GC systems, four seem to
have small orbital inclinations (i.e., with $\cos i\,>\,0.8$), the
exception being NGC~1851A. {\em A priori}, one would expect only one
out of five systems to have such a small $i$. We have used
a Kolmogorov-Smirnov test to compare the $\cos i$ values corresponding
to the nominal $\dot{\omega}$ values and $M_p = 1.44 M_{\sun}$ (only
possible to calculate for the five massive binaries) with a fake set of
100 randomly oriented binary systems (i.e., with a uniform
distribution of $\cos i$). We obtain a 0.46\% probability that the
observed distribution is extracted from this set with random
orbital orientations. If we use instead $M_p = 1.2 M_{\sun}$, we can
calculate $\cos i$ for all the eccentric GC binaries. The probability
that the resulting distribution of $\cos i$ is selected from the set
with random orientations is $7.7 \times 10^{-5}$.

Such low inclinations might be less unlikely were there a tendency
for pulsars to emit in a plane perpendicular to their orbit. During
accretion, orbital angular momentum is transferred to the NS,
making its rotation axis nearly perpendicular to its orbital
plane. If the angle between the magnetic and rotational axes of
pulsars ($\alpha$) is small, the magnetic axis will describe a narrow
cone nearly perpendicular to the orbital plane. However, no such
tendency for small $\alpha$ has been described in the literature.
If a pulsar has a small $\alpha$, its beam will probably illuminate
a smaller fraction of the sky, particularly if it is narrow. This
can only make the low-$\alpha$ objects {\em less} likely to be
detected\footnote{The pulse profiles also tend to be wider in these
nearly aligned rotators, further hindering their detection. We see
no correlation between the ``apparent'' inclinations in
Fig.~\ref{fig:binaries} and the pulse-widths, although the pulsar
profile of M5B is quite broad (Fig. \ref{fig:profiles}).}.
Furthermore, no such tendency towards low orbital inclinations is seen
among the two lighter binary MSPs in GCs, nor among the systems
with estimated orbital inclinations: NGC~6752A has $i > 70^\circ$
\cite{bkkv06}; PSR~J1909$-$3744 has $i = 86.6(1)^\circ$
\cite{jhb+05,hbo06}; PSR~J0437$-$4715 has $i \sim 43 ^\circ$
\cite{vbs+08} and the massive binary PSR~J1903+0327 has $i \sim 79
^\circ$ \cite{crl+08}\footnote{In many cases, high orbital inclinations
enable the measurement of the mass of a MSP through Shapiro delay,
producing a selection of such inclinations among the systems with
measured mass. This should not be too strong in the
sample we selected, systems with the timing precision of
PSR~J1909$-$3744 and PSR~J0437$-$4715 can have their Shapiro delays
measured almost irrespective of their orbital inclination}.

If the low mass functions of the massive binaries are not a result of
systematically low orbital inclinations, they can only be due to
systematically small companion masses. This implies that in the
majority of these systems the pulsar masses are significantly larger
than in the lighter binaries.

\begin{figure}
\epsscale{1.18}
\plotone{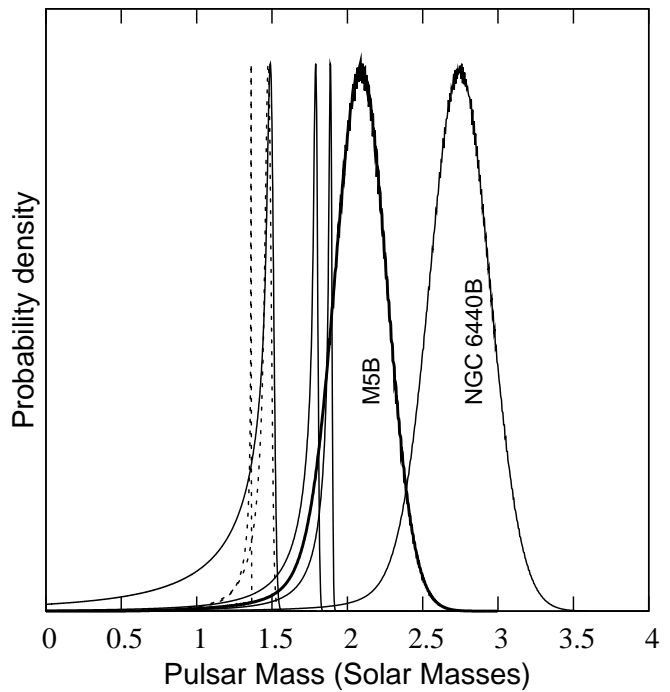}
\caption{\label{fig:psr_masses} Probability distribution functions
(pdfs) for the eccentric MSPs binaries in GCs (see also
Fig.~\ref{fig:binaries}). The mass pdfs of the MSPs in the less
massive binaries (those with $M < 2 M_{\sun}$) are represented by the
dashed curves. The distribution of masses is much broader than is
found for the components of DNS systems. Four out of a total
of seven systems seem to be significantly more massive than the most
massive NS in DNS systems, PSR~B1913+16.}
\end{figure}

\subsection{Implications for the equation of state of dense matter}
\label{sec:massives}

Because there is no physically plausible reason to assume that most
massive binaries have a small $i$ we now assume that the
probability density of $\cos i$ is constant. We use this to calculate the
mass pdfs from $\dot{\omega}$ as described in \S~\ref{sec:masses}
for all the pulsars in Fig.~\ref{fig:binaries}. The pdfs are displayed
graphically in Fig. \ref{fig:psr_masses}. If the probability that
$1.2\,M_{\sun}\,<\,M_p\,<\,1.44\,M_{\sun}$ is small
(as is the case for Terzan 5 I and J, M5B and NGC~6440B), it is a
direct indication that the required orbital inclination ranges are
very narrow (as shown in Fig.~\ref{fig:binaries}) and therefore
unlikely under the present assumption. For M5B, this probability is only
0.77\%, while for NGC~6440B it is even smaller, only 0.10\%.
Multiplying such probabilities for all the massive MSPs in GCs, we
obtain a composite probability of $5.3 \times 10^{-9}$ that all
have masses between 1.2 and $1.44\,M_{\sun}$. It is therefore very
likely that some of these NSs are significantly more massive.

As discussed above in \S \ref{sec:masses}, for M5B there is a 95\%
probability that the pulsar is more massive
than $1.72\,M_{\sun}$. This would exclude a third of the equations
of state considered in Lattimer \& Prakash (2007). However, in this
respect NGC~6440B should be far more constraining; there are 99 and 95\%
probabilities that the mass of that pulsar is $>$ 2.01 and
$2.36\,M_{\sun}$ respectively. We consider the M5B result to
be more secure: the non-detection of its companion (\S \ref{sec:optical})
almost guarantees that its $\dot{\omega}$ is purely relativistic (\S
\ref{sec:relativistic}).

There are two MSPs listed in Table~\ref{tab:msp_masses} for which
large masses have already been determined, PSR~J0437$-$4715 and
PSR~J1903+0327. The latter in particular 
has the potential for a precise, unambiguous measurement of a large
pulsar mass in the very near future. At the moment, it has not been
confirmed whether its $\dot{\omega}$ is relativistic or not, although
it is likely to be so. These results strengthen the case for the
existence of massive NSs.

\section{Formation of massive neutron stars}

The MSP mass estimates in Table~\ref{tab:msp_masses} and the
mass pdfs in Fig.~\ref{fig:psr_masses}, especially those of M5B and
NGC 6440B, suggest that the distribution of MSP masses could span a
factor of 2, a situation that is completely different to that found
for the components of DNSs. NGC~1851A and the MSPs in the ``light''
($M\,<\,2\,M_{\sun}$) binaries have masses smaller than
$1.5\,M_{\sun}$, i.e., they are not significantly more massive than
mildly recycled NSs, despite having spin frequencies of
hundreds of Hz. In particular, the case of M28C demonstrated that if
all NSs start with $M_P > 1.2 M_{\sun}$, then some MSPs
can be recycled by accreting $<\,0.15\,M_{\sun}$ from their companions.
At the other end of the distribution, NGC~6440B could be twice as
massive.

It could be that MSPs were born with this wide range of masses.
Hydrodynamical core collapse simulations \cite{tww96,bok+08} indicate
that stars below $\sim 18\,M_{\sun}$ form $\sim 1.20-1.35\,M_{\sun}$
NSs (such as 47~Tuc~H, M28C and NGC~1851A), while stars with
masses between $18 - 20\, M_{\sun}$ form $1.8\,M_{\sun}$ NSs (similar
to PSR~J0437$-$4715, PSR~J1903+0327, Terzan 5 I, J and M5B).
Above $20\,M_{\sun}$, stars experience partial fall-back of material
immediately after the supernova that can significantly increase the
mass of the stellar remnant, making it either a super-massive NS (like
NGC~6440B) or a black hole.

This possibility raises the question of why such massive NSs, while
representing about half of the MSPs in Table~\ref{tab:msp_masses}, have
not been found among the 9 known DNS systems. Most of the secondary NSs
in DNS systems have masses between 1.2 and $1.3\,M_{\sun}$, and recently
van den Heuvel (2007)\nocite{heu07} suggested that these were formed by
electron capture (EC) supernovae. The nine primary NSs
in DNS systems were still likely formed in normal (iron core
collapse) supernovae. The predicted percentage of massive NSs is quite
small, and with only 9 known DNS we are unlikely to see any massive NSs
as the primary \cite{bok+08}, a situation that is very different from
what is derived for the MSPs.

If the extra mass of some MSPs were instead acquired during the long
accretion episodes that recycled them,  we can explain naturally why
we only see massive NSs as MSPs (not just in
GCs, but also in the Galaxy, e.g. Verbiest et al. 2008, Champion
et al. 2008) but not in DNS systems.

\section{Conclusion}

We have measured the positions and proper motions of M5A and B.
This has allowed a detailed search for the companion of M5B. However,
no object was detected within $2.5 \sigma$ of the position of M5B
to a magnitude limit of 26-26.5, indicating that its companion is
either a low-mass MS star or a WD. We have measured the rate of
advance of periastron for this binary system, concluding that it is
very likely due solely to the effects of general relativity.  In this
case, the total mass of the binary is $2.29 \pm 0.17 M_{\sun}$,
similar to the total masses of Terzan~5~I and J. Like those pulsars
and NGC~6440B, the relatively low mass function for M5B indicates that
most of the system mass is likely to be in the pulsar, which we
estimate to be $2.08 \pm 0.19 \,M_{\sun}$
(1 $\sigma$). There is a 95\% probability that the mass of this pulsar
is above $1.72\,M_{\sun}$. If confirmed, this would exclude
about a third of the equations of state that are now accepted as possible
descriptions of the bulk properties of super-dense matter. 

Together with other recent results, the large mass derived for M5B
suggests that MSPs have a very broad mass distribution; half
of these objects seem to be significantly more massive than
$1.44\,M_{\sun}$. It is likely that all NSs began with the a narrow
mass range like that found in DNS. They then accretted different
amounts of matter (in some cases as much as their starting mass)
during their evolution to the MSP phase.

\acknowledgements

We thank S. M. Ransom and I. H. Stairs for many of the L-band 
observations made since 2001, Patrick Lazarus and Melissa Ilardo for
help with data reduction, Chris Salter, Marten van Kerkwijk and the
referee, Matthew Bailes, for their many constructive suggestions.
J.W.T.H. thanks NSERC and the Canadian Space Agency for a postdoctoral
fellowship and supplement respectively.  The Arecibo Observatory, a
facility of the National Astronomy and Ionosphere Center, is operated
by Cornell University under a cooperative agreement with the National
Science Foundation. The Borg was funded by a New Opportunities Grant
from the Canada Foundation for Innovation. This paper makes use of
data obtained from the Isaac Newton Group Archive which is maintained
as part of the CASU Astronomical Data Centre at the Institute of
Astronomy, Cambridge.

\typeout{get arXiv to do 4 passes: Label(s) may have changed. Rerun}

\end{document}